\documentclass[trackchanges,rnaas]{aastex701}
\usepackage{hyperref}

\begin{document}

\title{A Possible Protocluster of Galaxies Serendipitously Discovered in the Field of an Intermediate-Redshift Post-starburst Galaxy}

\author[]{Mary~C.~Knowlton}
\affiliation{Department of Physics and Astronomy, Texas A\&M University, 578 University Dr., College Station, TX, 77843, USA}
\email[]{maryknowlton@tamu.edu}  

\author[0000-0003-3256-5615]{Justin~S.~Spilker}
\affiliation{Department of Physics and Astronomy, Texas A\&M University, 578 University Dr., College Station, TX, 77843, USA}
\email[]{jspilker@tamu.edu}

\author[0000-0001-5063-8254]{Rachel~Bezanson}
\affiliation{Department of Physics and Astronomy and PITT PACC, University of Pittsburgh, Pittsburgh, PA 15260, USA}
\email[]{rachel.bezanson@pitt.edu}

\author[0000-0002-1759-6205]{Vincenzo~R.~D’Onofrio}
\affiliation{Department of Physics and Astronomy, Texas A\&M University, 578 University Dr., College Station, TX, 77843, USA}
\email[]{donofr19@tamu.edu}

\author[0009-0005-4226-0964]{Anika~Kumar}
\affiliation{Laboratory for Multiwavelength Astrophysics, School of Physics and Astronomy, Rochester Institute of Technology, 84 Lomb Memorial Drive, Rochester, NY 14623, USA}
\email[]{ak8532@rit.edu}

\author[0000-0003-4075-7393]{David~J.~Setton}
\altaffiliation{Brinson Prize Fellow}
\affiliation{Department of Astrophysical Sciences, Princeton University, Princeton, NJ 08544, USA}
\email[]{davidsetton@princeton.edu}

\author[0000-0002-1714-1905]{Katherine~A.~Suess}
\affiliation{Department for Astrophysical \& Planetary Science, University of Colorado, Boulder, CO 80309, USA }
\email[]{suess@colorado.edu}

\begin{abstract}
We present the serendipitous discovery of an overdensity of submillimeter galaxies (SMGs) in the field of SDSS~J0909-0108, a massive $z\sim0.7$ post-starburst galaxy from the SQuIGG$\vec{L}$E survey. ALMA observations at 870\,$\mu$m and 2\,mm reveal six galaxies within a 35'' region with flux ratios consistent with emission from dust. Given the rarity of 870\,$\mu$m sources and the small field-of-view of ALMA, we speculate that some of these sources are physically associated. None of the sources are at the same redshift as the post-starburst, and four do not have spectroscopic redshifts. We suggest that follow-up optical and/or ALMA observations be carried out to measure redshifts for the galaxies in this potential protocluster environment.

\end{abstract}


\section{Introduction} \label{sec:intro}
The massive galaxy clusters of the present day Universe evolved from much more diffuse collections of galaxies in the early Universe; we call these progenitors ``protoclusters" \citep{Overzier_2016}. Protoclusters are expected to form most of their stars at $z>2$ in galaxies that individually form stars at high rates \citep{Papovich_2010}.

Protoclusters can be identified in many ways, including as overdensities of rare galaxy types such as quasars or  highly luminous starburst galaxies, which can be observed at radio and submillimeter wavelengths at high redshifts \citep{Miley_HzRGs}. Many protoclusters have been found to contain a strong overdensity of dusty, star-forming galaxies, including well-known examples like the Spiderweb ($z=2.16$; \citealt{Dannerbauer_2014}), SSA22 ($z=3.09$; \citealt{Umehata_2015}), the Distant Red Core (DRC, $z=4.00$; \citealt{Oteo_2018,Long_2020}), and SPT2349-56 ($z=4.30$; \citealt{Miller_2018,Hill_2020}). These latter two were first identified as candidate protoclusters on the basis of the dusty galaxies they contain, selected from wide-area far-IR or millimeter surveys. 

Here, we present the serendipitous discovery of an overdensity of sources found in a single ALMA pointing at both 870 $\mu$m and 2 mm. Due to the rarity of submm-bright sources, we speculate that the continuum-detected sources may be physically associated at the same redshift, implying they are potentially members of a protocluster discovered by pure chance. This Research Note presents the discovery of these sources and a first attempt at determining their redshifts.



\section{Data and Methods} \label{sec:methods}
This work makes use of ALMA Band 4 (2\,mm) and Band 7 (870\,$\mu$m) observations, published in \citet{Setton_2025} and \citet{donofrio_2026}, respectively. The single-pointing observations were designed to detect the CO(2--1) and CO(5--4) emission lines in SDSS~J0909-0108, a massive post-starburst galaxy at $z=0.7021$ selected from the SQuIGG$\vec{L}$E survey \citep{Suess_2022}. In Band~7, we combined observations from projects 2024.1.00216.S (PI: J. Spilker) and 2024.1.01252.S (PI: V. D'Onofrio). Further details on the ALMA observational design are available in the aforementioned works.

We used CASA \citep{CASA} to create continuum images at both wavelengths using natural weighting of the visibilities. These images, not corrected for the primary beam response, are shown in Fig.~\ref{fig:fig1}. We also created spectral data cubes of the lower and upper sidebands of both datasets to search for spectral lines with cube channel widths of 100\,km/s. We measured primary-beam-corrected continuum fluxes and searched visually for spectral lines using the CARTA viewer \citep{CARTA}; we additionally used the LineSeeker software\footnote{\url{https://github.com/jigonzal/LineSeeker}\label{lineseeker}} to search for candidate emission lines. We also compared the ALMA sub/millimeter images to optical images of the SDSS~J0909-0108 field from SDSS SkyServer-Navigate \citep{SDSS_SkyServer} and the Hyper Suprime-Cam (HSC) \citep{10.1093/pasj/psab122}.


\section{Results and Discussion} \label{sec:results}
We identified six continuum-emitting sources in the field surrounding SDSS~J0909-0108, which are labeled in the top panels of Fig.~\ref{fig:fig1} with spectra shown below. For all six sources, the $S_{870\mu\mathrm{m}}$/$S_{2\mathrm{mm}}$ flux ratios indicate steeply-rising SEDs consistent with emission from cold dust.

We rule out that these sources are at the same redshift as the post-starburst galaxy -- while both CO(2--1) and CO(5--4) are clearly detected at 135.44 and 338.56\,GHz in SDSS~J0909-0108 itself \citep{donofrio_2026}, lines at these frequencies are not detected in any of the other sources, and no CO(2–1) emitters are found nearby \citep{Kumar_2025}. In source~6, we found candidate spectral lines at 138.123 and 147.441\,GHz, which we identify as CO(4--3) and [CI](1--0) at $z=2.338$. Source~2 is an optically-bright spiral galaxy in the SDSS and HSC imaging, which we identify as GAMA~3909836 at $z=0.258$. We did not find any lines in the other four sources, implying that they are not at $z = 0.702$ or 2.338. We also rule out many small redshift ranges based on the nondetection of lines, but large swaths of redshifts remain viable.

The middle panels of Fig.~\ref{fig:fig1} show cutouts of the HSC i-band image overlaid with ALMA contours. The four sources that do not have known redshifts are either optically faint (1, 4, 5) or nondetections (3). Several sources are near other much brighter sources, which makes their HSC catalog photometry unreliable. Estimating photometric redshifts is beyond the scope of this Note due to the sources' proximity to these brighter objects.

The brightest source, source~1, has $S_{870\mu\mathrm{m}} = 8.4$\,mJy. Such bright submm sources are exceedingly rare, with sky densities $N(S_{870\mu\mathrm{m}} > 8.4\mathrm{mJy}) \approx 35$\,deg$^{-2}$ \citep{SCUBA_numbercounts}. We have observed $\approx$50 SQuIGG$\vec{L}$E post-starbursts with ALMA at 2\,mm, covering a total sky area $\approx10^{-4}$\,deg$^2$, implying a $\lesssim$0.5\% chance to find even a single submm source as bright as source~1 in our survey area.

The fact that source~1 is an extremely bright 870\,$\mu$m source surrounded by (at least) three other sources with $S_{870\mu\mathrm{m}} \sim 1.5-2.5$\,mJy leads us to hypothesize that these four galaxies may be physically associated, i.e. may be members of an overdense galaxy protocluster. Alternatively, they may be simply chance projections along the line of sight, exacerbated by the strongly-negative K-correction at submm wavelengths that results in sources at many redshifts appearing with similar flux density \citep[e.g.][]{Blain_2002,Casey_2014}. 

If these four dusty sources are physically associated, their fluxes and sky separations are not dissimilar to the central `core' regions of other high-redshift protoclusters like SPT2349-56 or the DRC \citep{Hill_2020,Long_2020}. If this overdensity represents the core of a galaxy protocluster, source 1 could be its central star-forming galaxy. A further spectral line survey of this region should be conducted to determine whether these sources do, in fact, represent an early protocluster of galaxies at the same redshift. Given the faintness of these sources at optical wavelengths, we suggest that a natural next step would be an ALMA spectral scan to determine redshifts based on CO and/or CI lines.

\begin{figure*}[]
    \begin{center}
        \includegraphics[width=1\linewidth]{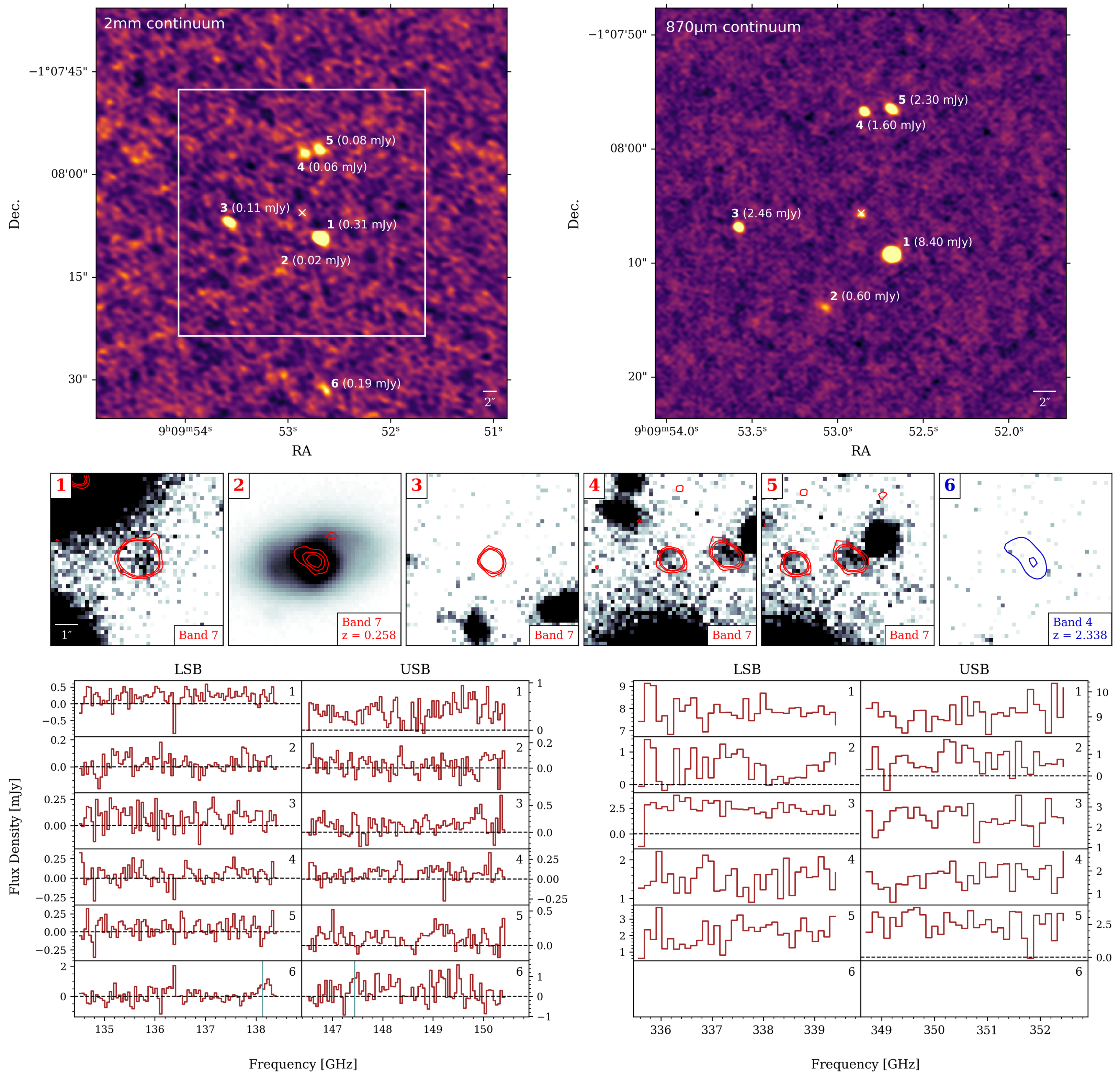}
    \end{center}
    \caption{Top left: ALMA Band 4 continuum image. The white box indicates the FOV of the Band 7 image. All six sources are labeled with their primary-beam-corrected fluxes in parentheses. Top right: ALMA Band 7 continuum image, with primary-beam-corrected flux densities in parentheses for the five visible sources. The location of the SDSS~J0909-0108 post-starburst galaxy ($z=0.7021$) is marked by an X in both images. Middle: HSC i-band cutouts of each source overlaid with ALMA Band 7 (red) and Band 4 (blue) contours at 3$\sigma$, 5$\sigma$, and 7$\sigma$. Lower left: spectra for each source in the Band 4 upper and lower sidebands. The blue vertical lines mark the frequencies of the two line candidates found in source 6. Lower right: spectra for each source in the Band 7 upper and lower sidebands.}
    \label{fig:fig1}
\end{figure*}





\begin{acknowledgments}
MCK and JSS thank the TAMU College of Arts \& Sciences Undergraduate Research Program for their support. JSS, VRD, and KAS acknowledge support from the National Science Foundation under NSF-AAG Nos. 2407954 \& 2407955. This research note makes use of the following ALMA data: ADS/JAO.ALMA\#2021.1.01535.S, ADS/JAO.ALMA\#2024.1.00216.S, ADS/JAO.ALMA\#2024.1.01252.S. ALMA is a partnership of ESO (representing its member states), NSF (USA) and NINS (Japan), together with NRC (Canada), NSTC and ASIAA (Taiwan), and KASI (Republic of Korea), in cooperation with the Republic of Chile. The Joint ALMA Observatory is operated by ESO, AUI/NRAO and NAOJ. The National Radio Astronomy Observatory and Green Bank Observatory are facilities of the U.S. National Science Foundation operated under cooperative agreement by Associated Universities, Inc.
\end{acknowledgments}

%
\facility{ALMA, Subaru/HSC}

\software{CASA \citep{CASA}, CARTA \citep{CARTA}, \texttt{astropy} \citep{Price-Whelan_2018}, \texttt{matplotlib} \citep{matplotlib}, LineSeeker\footref{lineseeker}}



\bibliography{ref}{}
\bibliographystyle{aasjournalv7}



\end{document}